\def\choose#1#2{\left(\begin{array}{c} #1 \\ #2 \end{array}\right)}

\def\emla{e^{-\lambda t_0}}

\def\Var{\mbox{Var}}

\def\M3{M_{3+}}

\def\Var{\mbox{Var}}

\documentclass{statsoc}
\usepackage{epsfig}
\usepackage{epic,graphics}

\title[Boolean models for particle flow experiments]{M-estimation of Boolean models for particle flow experiments}
\author{Jason A. Osborne}
\address{Department of Statistics, North Carolina State University,
Raleigh, NC,
USA}
\author[J.A. Osborne and T.E. Grift]{Tony E. Grift}
\address{Department of Agricultural and Biological Engineering, University of Illinois,
Urbana, IL,
USA}
\begin{document}
\begin{abstract}
Probability models are proposed for 
passage time data collected in experiments with
a device designed to measure particle flow during
aerial application of fertilizer. 
Maximum likelihood estimation of flow intensity is
reviewed for the simple linear Boolean model,
which arises with the assumption that each particle
requires the same known passage time.  M-estimation is
developed for a generalization of the model in which
passage times behave as a random sample from a
distribution with a known mean.  The generalized
model improves fit in these experiments.
An estimator of total particle flow is constructed by
conditioning on lengths of multi-particle clumps.
\end{abstract}

\section{Introduction}
Measuring the outflow of granular particles from an airborne spreader 
during the aerial application of fertilizer or pesticide presents 
agricultural engineers with a difficult problem.  
The goal of uniform distribution over a targeted field requires 
knowledge about flow rate of the material as it is dropped from
the aircraft.
Windspeed, air speed, 
granule properties, humidity, and temperature 
have been identified (\cite{casady}) as factors which can lead to 
variability in these outflow rates and hence amounts of material that 
reach the target.   
Typically, applicators are calibrated annually 
so that they achieve an average target flow rate.  
In practice, pilots use a simple lever-operated gate to change the flow rate
in order to account for extreme values of these factors.
This adjustment is based on 
intuition, without any feedback from measurement of particle flow.

One approach to providing the pilot with more information uses
an optical sensor device (\cite{grift4}) which measures the velocity 
(in meters per second)
and size of clumps
of particles as they flow through the spreader duct. 
%
This device has two photo-sensitive arrays of optical sensors that 
receive a signal from a light source.  
As a particle passes an
active area, 
it blocks this light thereby interrupting the signal received by the
sensors.
As long as 
all of the sensors in the 
array 
are receiving a high signal, the channel is 
classified as
unoccupied and
this is taken as an indication that there are no particles
flowing through at that instant.
If the signal to any one of sensors is interrupted,  
this is interpreted as the presence of 
at least one particle, constituting a clump, in flow.  
The two sensor arrays are $0.00078$ meters apart and it is possible
to measure the time in seconds that it takes a clump to move from one array
to the other, $\Delta t_f$.
The total time that either array is blocked,
$\Delta t_b$, is also measured, facilitating 
calculation of velocity
in meters per second,
$v=0.00078/\Delta t_f$ 
and clump length in meters, $ CL = v \Delta t_b.$
These observable clump lengths, either in terms of physical length in meters 
or time in seconds, are 
the basis for inference about
particle flow in the system.
Such a measurement device is called a 
type II counter (\cite{pyke1}).  

\citet{grift2} and \citet{grift3} carried out bench-scale experiments to 
evaluate the optical sensor device in situations designed to simulate
the flow of fertilizer particles through an airborne spreader duct.
In these experiments, a known number 
of spherical particles with a known mean diameter of $4.45mm$ was dropped 
from predetermined heights through a duct on which the sensor
device was installed.
The heights from which the particles were dropped was controlled at
several values to simulate a range of particle velocities and flow rates.
A histogram representing the distribution of particle clump lengths
in units of time, obtained from one run of these experiments, is 
shown in Figure 1.
The relative frequencies for clump lengths (in $msec$) are based on dropping
4000 spherical steel particles (actually BBs) from a fixed height.



\begin{figure}
\centering
\makebox{\includegraphics[height=2.9in, width=3in]{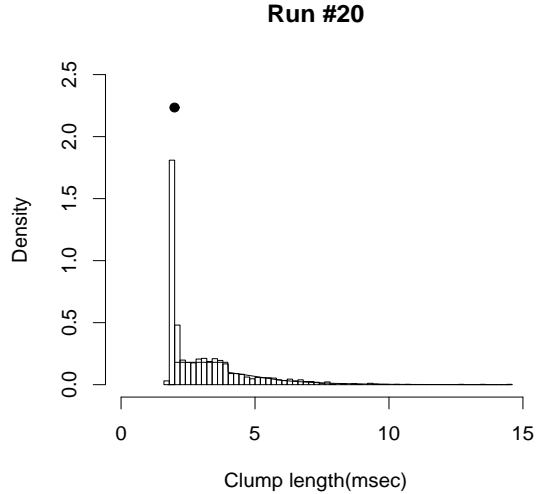}}
\caption{Probability histogram of $N(t)=1790$ clump lengths in $msec$}
\end{figure}

In this paper, simple linear Boolean models are used to describe
the clump length data generated by the Grift experiments,
thus providing a basis for inference about flow rates and total particle
flow during a dispersal period.
In particular, flow intensity is quantified 
by a single rate parameter in a simple Boolean model.  
Maximum-likelihood is 
reviewed 
in cases where particles require a fixed time for passage
and an $M$-estimator is obtained in more general cases.  
Assessment of total particle flow utilizing this estimator
is also developed.

Section 2 introduces the Boolean models and establishes notation
and terminology.  Results for the clump length distribution 
derived in \citet{hall1} are used to develop $M$-estimation
of flow intensity
and the $M$-estimator is compared with maximum likelihood and
other moment estimators by simulation.  
In section 3, two estimators of total particle flow are proposed,
including one obtained by derivation of the conditional expectation of 
the number of particles in a clump
given clump length under the equal diameters model.  
Simulations are carried out to give an assessment of the performance
of this estimator in the random passage times model.
The methods are evaluated based on their performance with the 
experimental data in Section 4.  
Section 5 concludes.


\section{Estimation}

To obtain a probability model for the clump-length data, particles are assumed 
to be identically spherical with a known diameter, $d_0$,
and to arrive at the sensor according to a 
homogeneous Poisson process with unknown intensity $\lambda.$   
Passage of particles is assumed to continue unabated upon arrival at the sensor.  
In one version of the model, the particles are travelling at a constant 
velocity, say $v_0$,
and the {\em segment length} (\cite*{hall1}), or time required for any 
single particle to pass the
sensor, is constant at $t_0=d_0/v_0$.   
In a second version, velocities or diameters 
are assumed to vary in such a way that segment lengths behave as a 
random sample from a population with a known mean $\mu$
and an unknown variance $\sigma^2$.
The two models will be referred to as
deterministic segment length (DSL) or random segment length (RSL) 
models, respectively.

Suppose that particle flow is observed for $t$ time units.
Let the number of particles arriving at the sensor in this time
period be denoted by $A(t)$.
Let $N(t)$ denote the number of complete
particle clumps observed by time $t$.  Let 
$Y_1,Y_2,\ldots,Y_{N(t)}$ denote 
the lengths of these clumps 
and $Z_1,Z_2,\ldots$ the spacings between them.
Let the unobservable number of particles comprising clump $i$ be called
the clump {\em order} and be denoted by $K_i$.

Figure 2 illustrates the clumping process using an example with $A(t)=7$ 
particles arriving at a sensor 
at times $1.9, 5.9,6.8,7.5,11.6,12.8$ and $17.1 \ msec$
during an observation period of $t=20 \ msec$.
If particles are assumed to have diameter $4.45\ mm$ and to be travelling at a 
constant velocity of $2.225 \ mm/msec$, the passage time required for each,
or deterministic segment length, is $d_0=2 \ msec$, leading to four 
clumps of lengths 
$y_1=2,y_2=3.6,y_3=3.2,y_4=2 \ msec $ that exit the sensor at times
$3.9,9.5,14.8$ and $19.1 \ msec$, respectively.  Spacings between clumps
would be of length $z_1=1.9,z_2=2.0, z_3=2.1$ and $z_4=2.3\ msec$ 
and the four 
clump orders would be 
$k_1=1,k_2=3,k_3=2,k_4=1$.

\setlength{\unitlength}{4mm}
\begin{figure}
\begin{center}
\begin{picture}(20,22)(-2,-2)
\linethickness{1pt}
\put(0,18){\vector(1,0){20}}                       
\put(-5,18){Arrival}
\put(-5,17){times}
\thicklines
\dashline{0.4}(1.8,18)(1.8,16)
\dashline{0.4}(5.9,18)(5.7,16)
\dashline{0.4}(6.8,18)(6.8,14)
\dashline{0.4}(7.5,18)(7.5,12)
\dashline{0.4}(11.6,18)(11.6,16)
\dashline{0.4}(12.8,18)(12.8,14) 
\dashline{0.4}(17.1,18)(17.1,16) 
\put(1.8,16){\vector(1,0){2}}
\put(5.7,16){\vector(1,0){2}}
\put(6.8,14){\vector(1,0){2}}
\put(7.5,12){\vector(1,0){2}}
\put(11.6,16){\vector(1,0){2}}
\put(12.8,14){\vector(1,0){2}} 
\put(17.1,16){\vector(1,0){2}} 
\put(1.8,18){\circle*{0.8}}
\put(5.9,18){\circle*{0.8}}
\put(6.8,18){\circle*{0.8}}
\put(7.5,18){\circle*{0.8}}
\put(11.6,18){\circle*{0.8}}
\put(12.8,18){\circle*{0.8}}
\put(17.1,18){\circle*{0.8}}
\put(0,5){\vector(1,0){20}}                     
\put(0,5){\vector(0,1){6}}                      
\multiput(0,5)(0,1){6}{\line(-1,0){.3}}            
\multiputlist(-0.7,5)(0,1){0,1,2,3,4}    
\put(1.8,5){\line(0,1){1}}   
\put(1.8,6){\line(1,0){2}}   
\put(3.8,5){\line(0,1){1}}   
\put(5.9,5){\line(0,1){1}}  
\put(5.9,6){\line(1,0){0.9}}
\put(6.8,7){\line(1,0){0.7}}    
\put(6.8,6){\line(0,1){1}}    
\put(7.5,7){\line(0,1){1}}  
\put(7.5,8){\line(1,0){0.4}}
\put(7.9,7){\line(0,1){1}}    
\put(7.9,7){\line(1,0){0.9}}    
\put(8.8,6){\line(0,1){1}}    
\put(8.8,6){\line(1,0){0.7}}    
\put(9.5,5){\line(0,1){1}}    
\put(11.6,5){\line(0,1){1}}  
\put(11.6,6){\line(1,0){1.2}}  
\put(12.8,6){\line(0,1){1}}   
\put(12.8,7){\line(1,0){0.8}} 
\put(13.6,6){\line(0,1){1}} 
\put(13.6,6){\line(1,0){1.2}} 
\put(14.8,5){\line(0,1){1}} 

\put(17.1,5){\line(0,1){1}}   
\put(17.1,6){\line(1,0){2}}   
\put(19.1,5){\line(0,1){1}}   

\put(-5,9){Particles}
\put(-5,8){in duct}
\put(0,0){\vector(1,0){20}}                     
\multiput(0,0)(5,0){4}{\line(0,1){.3}}         
\put(-5,1){Clump-}
\put(-5,0){lengths}
\multiputlist(0,-0.5)(5,0){0,5,10,15,20}    
\put(1.8,1){\vector(1,0){2}}
\put(3.8,1){\vector(-1,0){2}}
\put(2.7,1.5){$y_1$}
\put(5.9,1){\vector(1,0){3.6}}
\put(7.5,1.5){$y_2$}
\put(9.5,1){\vector(-1,0){3.6}}
\put(11.6,1){\vector(1,0){3.2}}
\put(13.0,1.5){$y_3$}
\put(14.8,1){\vector(-1,0){3.2}}
\put(17.1,1){\vector(1,0){2}}
\put(17.9,1.5){$y_4$}
\put(19.1,1){\vector(-1,0){2}}
\put(9,-2){Time}
\end{picture}
\end{center}
\caption{A diagram of particle flow measurement using a type II counter}
\end{figure}
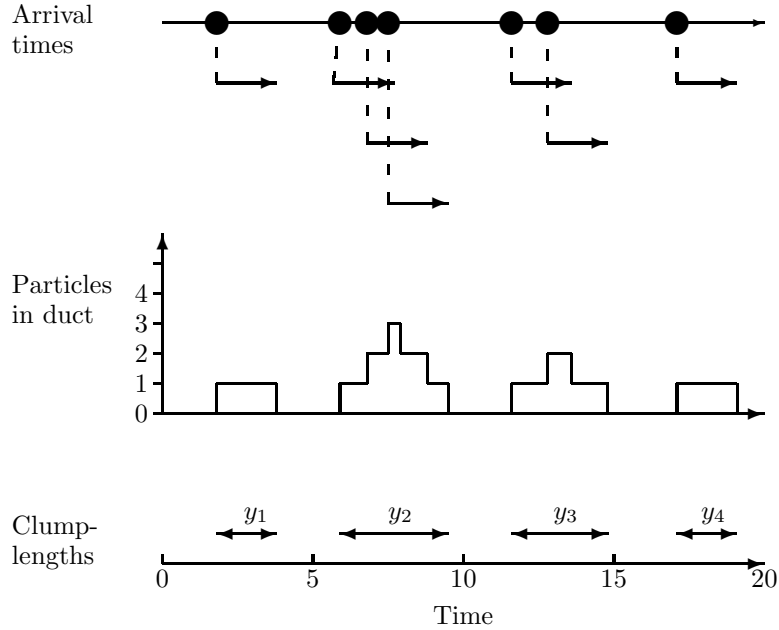


%


The particle clumps constitute a coverage process on one dimension.
\cite{hall1} describes the process as a simple linear Boolean model;
simple because the clump-lengths are line segments and linear
%
%
because the events occur in one dimension, the time line.
Linear Boolean models also arise as linear transects from higher
dimensional convex-grain Boolean models.  
In the language of queueing theory, the 
number of particles in a clump at the sensor at a given time forms 
an $M/D/\infty$ queue in the DSL model and
an $M/G/\infty$ queue in the RSL model 
and clump-lengths are called {\em busy periods}.  
There is much literature on these models from queueing theory
(\cite*{daley}).  
For statistical inference for
the distribution of diameters or more complex quantities 
describing the grain process, or for Boolean models 
in higher dimensions, see \cite{molchanov}.
\cite{handley2} derived a discrete approximation to the
distribution of clump-length in the linear Boolean model and
used it for likelihood inference. 
\cite{crespi} have employed the linear Boolean model for monitoring events of
viral activity in human subjects.

\subsection{Likelihood}
Specification of the clump-length density, $f(y;\lambda)$ is 
difficult
outside of the case where particle diameters are degenerate.
In the DSL model, \cite{hall1} has shown that the density 
has point mass $\emla$ at $y=t_0$, and is otherwise 
given by
\begin{eqnarray*}
\lefteqn{f(y;\lambda,t_0)=\lambda \frac{\emla}{1-\emla}}\\
 &   &\left[1+\sum_{j=1}^{s-1} \frac{(-1)^j}{j!}\{\lambda(y-(j+1)t_0)\}^{j-1}
e^{-j\lambda t_0}\{\lambda(y-(j+1)t_0)+j\}\right]
\end{eqnarray*}
where $y>t_0$ and $s$ is the largest integer such that $t_0<y/s$.
The continuous part of the density is uniform over $(t_0, 2t_0)$, 
and  decreasing 
for $y>2t_0$. 
For small $\lambda$, $f(y;\lambda,t_0)$
can be approximated by the uniform distribution on $(t_0,2t_0)$, for 
large $\lambda$ it can be approximated by an 
exponential distribution. 
The fitted density $f(y;\lambda=0.40,t_0=2.00 msec)$ overlays the probability 
histogram of experimental clump-lengths in Figure 1.


For RSL models, likelihood inference is difficult because of the
complexity of the clump-length distributions (\cite*{handley1}).
In the DSL model, an approximate likelihood function can be 
specified by ignoring the residual lifetime of the process.  The 
residual lifetime is the duration of the last incomplete clump or 
spacing.  A clump is a singleton if there are no arrivals within $t_0$ 
time units of the start of the clump, an event which occurs with probability
$\emla$.  Let $M_1=\#\{y_i:y_i=t_0\}$ denote the number of singleton clumps.
By independence of clump-lengths, the approximate partial Boolean 
likelihood can be factored into components for singleton point 
masses and multi-particle clump-length densities:

\begin{eqnarray*}
\tilde{\cal{L}}(\lambda;y_1,\ldots,y_{N(t)}) &=&
\underbrace{e^{-m_1\lambda t_0}}_{\mbox{singletons}} 
\underbrace{\prod_{i:y_i>t_0} f(y_i;\lambda,t_0)}_{\mbox{multi-particle lengths.}}
\end{eqnarray*}
Spacings $z_1,z_2,\ldots$ 
are not available for the experiments analyzed in section 4.
For cases where the $z_i$ are available,
an approximate complete Boolean likelihood may be obtained by
multiplying the partial likelihood by the likelihood from an 
exponential random sample, $\lambda^N e^{-\lambda \sum z_i}$.
%


For large $t$, the maximum likelihood estimator of $\lambda$
based on $\tilde{\cal{L}}$
is approximately normally distributed.
However, the analytic expression for the Fisher information
is unwieldy, particularly for large clump-lengths, where the degree of the
polynomial components of the clump-length density is high.  Alternatively,
approximate confidence regions can be constructed from the likelihood ratio 
test statistic, which has an approximate $\chi^2$ distribution on 1 degree
of freedom.

In the RSL model, where segment lengths are distributed as a random
sample from a known distribution with distribution function $G(x;\theta)$, 
the clump-length density and resulting likelihood are considerably 
more complex.  Let $f_{RSL}(y;\lambda,\theta)$ denote the clump-length density,
which depends on the unknown parameters, $\lambda$ and $\theta$.
Ignoring the residual lifetime, the partial likelihood of the complete 
clumps is then
\begin{eqnarray*}
\tilde{\cal{L}}_{RSL}(\lambda,\theta;y_1,\ldots,y_{N(t)}) &=&
\lambda^{N} e^{-\lambda \sum z_i}
\prod_{i=1}^{N(t)} f_{RSL}(y_i;\lambda,\theta).
\end{eqnarray*}
\cite{hall1}
shows that the Laplace transform $\gamma$ of $Y$ is
$$ \gamma(s) = 1+ \frac{s}{\lambda}-
\left(\lambda\int_0^\infty \exp\{-st - \lambda \int_0^t\{1-G(x;\theta)\}dx\}dt\right)^{-1}.$$
\cite{stadje} obtains the clump-length distribution function $F_{RSL}(y)$ 
by inversion of $\gamma$, but it is an 
infinite sum of self-convolutions of a function that may 
involve an integral with no analytic solution, 
making inference 
based on $\tilde{\cal{L}}_{RSL}$ difficult.

\subsection{M-estimation}

An important issue in estimation of $\lambda$ is robustness under
model misspecification.  
Inspection of the clump-lengths from the experimental data, such as
the run depicted in Figure 1, reveals
that the number of clumps with lengths slightly in excess of $t_0$
is greater than expected, so that the distribution between $t_0$
and $2t_0$ is not uniform.  
This 
can be caused by
variability in diameter or velocity or by errors of measurement.  
A desirable property for any estimator 
is robustness to this departure from model assumptions.
%

For mean segment length $\mu$, the mean 
clump-length is given by
$$ E(Y;\lambda) = \frac{e^{\lambda \mu}-1}{\lambda}$$
in either the DSL or RSL model, regardless of the distribution 
of segment lengths (\cite*{hall1}).
For known $\mu$, consider the $M-$estimator $\tilde{\lambda}$
which satisfies
$$\bar{y} = \frac{e^{\tilde\lambda \mu}-1}{\tilde\lambda}.$$
A solution exists 
by the mean value theorem with $E(Y;\lambda)$ increasing 
in $\lambda$.  
Though there is no analytic solution, the equation
can be solved rapidly using any root-finding procedure, 
such as the {\tt uniroot} function in 
the $R$ statistical software package (\cite{rdoc}).
A starting point that works in simulations
is given by $\tilde\lambda=(\bar{y}-\mu)/(2\mu^2)$, which is
the solution obtained using a second order expansion 
of $e^{\tilde\lambda \mu}$ about 0.  An interesting aspect of
the sampling distribution of $\tilde\lambda$ is that it is negative
whenever $\bar{y} < \mu$, an event whose probability is small as
long as $\lambda \mu$ is not too small.

This estimating equation for $\lambda$ can be written
$$ \sum_i \psi(y_i,\lambda) =0$$
where $\psi(y,\lambda)=y-\lambda^{-1}(e^{\lambda \delta}-1)$.
Large-sample theory for $M$-estimators, 
(see, e.g. \cite*{boos})
can be used
for inference about $\lambda$.  
For a random
sample of $n$ clump-lengths $y_1,\ldots,y_n$, 
the asymptotic distribution of $\tilde\lambda$ is given by
$$ \sqrt{n}(\tilde\lambda-\lambda) \stackrel{\cal{L}}{\longrightarrow} N(0,C/B^2)$$
where $B$ and $C$ are functions of $\lambda$ defined by
\begin{eqnarray*}
B(\lambda)  & = & E(-\frac{\partial}{\partial \lambda}\psi(Y_1,\lambda))\\
C(\lambda)  & = & E( \psi^2(Y_1,\lambda)).
\end{eqnarray*}
Since $\psi$ is linear in $Y$, the expectation operations are
straightforward:
\begin{eqnarray*}
B(\lambda)  & = & \frac{e^{\lambda\mu}(\lambda \mu - 1) + 1}{\lambda^2} \\
C(\lambda)  & = & \Var(Y;\lambda).
\end{eqnarray*}
The variance of $Y$ depends on the distribution of segment lengths.
In the DSL model with $t_0=\mu$, 
$$ \Var(Y) = \lambda^{-2} (e^{2 \lambda \mu}-2\lambda \mu e^{\lambda \mu} - 1).$$
In the RSL model with segment lengths distributed according
to the general distribution function $G(x)$, 
clump-lengths have variance
$$ \Var(Y) = 2 \lambda^{-1} e^{\lambda \mu}\int_0^\infty \left(\mbox{exp}\left[\lambda 
\int_t^\infty (1-G(x)) dx\right]-1\right)dt-\lambda^{-2}(e^{\lambda \mu}-1)^2$$
which can be estimated using the sample variance
of clump-lengths, $s_y^2.$ 
Estimators for the variance of $\tilde\lambda$
are then given by 
$$ \widehat{\Var}(\tilde\lambda) = n^{-1}\frac{\tilde\lambda^2(e^{2\tilde\lambda \mu}-2 \tilde\lambda \mu e^{\tilde\lambda \mu} -1)}{(e^{\tilde\lambda\mu}(\tilde\lambda \mu -1)+1)^2}$$
in the DSL model and
$$ \widehat{\Var}_{\mbox{G}}(\tilde\lambda) = n^{-1}\frac{\tilde\lambda^4 s_y^2}{(e^{\tilde\lambda \mu}(\tilde\lambda \mu -1)+1)^2} $$
in either the DSL or RSL model. 
In large samples, approximate confidence intervals for $\lambda$ can be 
constructed 
from these 
estimates along with the normal approximation 
for $\tilde\lambda$.

\subsection{Other estimators}

For the DSL model with common deterministic passage time $t_0$, other 
method-of-moments (MOM) estimators can be constructed 
using only the clumpcount ($N(t)$) and singleton count ($M_1$) statistics.
The sequence of i.i.d. sums $\{Z_i+Y_i\}$ is a renewal process. 
Elementary renewal theory 
(\cite*{cox})
yields that as $t \rightarrow \infty$,
$$\frac{N(t)-t/\mu_R}{\sigma_R\sqrt{t/\mu_R^3}} \stackrel{\cal{L}}{\longrightarrow} N(0,1)$$
where $\mu_R$ and $\sigma_R^2$ denote the mean and variance of a randomly sampled
renewal period.  In DSL model with deterministic common passage time $t_0$,
\[
\begin{array}{ccccc}
\mu_R &=& E(Z+Y) &=&\lambda^{-1}e^{\lambda t_0} \\
\sigma_R^2 &=& \Var(Z+Y) &=&\lambda^{-2}(e^{2\lambda t_0}-2\lambda t_0 e^{\lambda t_0}).
\end{array} 
\]
Moments for $N(t)$ are then 
\begin{eqnarray*}
E[N(t)] & \approx & \lambda t e^{-\lambda t_0} \\
\Var[N(t)] & \approx & \lambda t \left(e^{-\lambda t_0} - 2\lambda t_0 e^{-2\lambda t_0}\right).
\end{eqnarray*}

The probability that a randomly 
selected clump is a singleton is $e^{-\lambda t_0}$ so that
$E(M_1)=\lambda t e^{-2\lambda t_0}$.  
A MOM estimator based on the observed number of singletons
is then
$$\tilde\lambda_{S}=-\frac{1}{t_0}\log\left(\frac{M_1}{N(t)}\right).$$
\cite{grift2} and \cite{grift3} base estimation of total mass flow 
on this estimator. 
%
Other 
estimators of $\lambda$ can be 
constructed by consideration of {\em vacancy}, $V\approx \sum Z_i$, 
or total time that that the sensor is unoccupied.  
\cite{hall1} develops asymptotic theory for a number of vacancy-based
estimators.
Measurements of $V$ were not available from the experiments discussed 
in section 4, and vacancy-based estimators are not considered further.

\subsection{Simulation} 


Simulations were undertaken to provide some information about 
the performance of these estimators, with three goals in particular:  
a comparison of the efficiency of the moment estimator $\tilde\lambda$ 
relative to the MLE
under the DSL model, an investigation of the robustness of the MLE under 
the RSL model and a comparison of coverage probabilities of confidence
intervals resulting from the two variance estimates of the asymptotically
normal $M$-estimator, $\tilde\lambda$.
Particle arrivals were generated according to a Poisson process.  
Three cases with an increasing degree of clumping 
were simulated using 
flow intensities of 
$\lambda=0.1,0.2$ and $0.3$.  Two times were considered for the length
of the total observation period, $t=1000$ and $t=10000$.
Preliminary experiments with particles far enough apart so that there
was no clumping indicated that measured passage times were normally
distributed.  So, passage times for individual particles were generated from a
normal distribution with 
a mean of $\mu=5$ with three different 
standard deviations, $\sigma=0,0.5,1$. 
The first of these standard deviations leads to the DSL 
model, the others to RSL 
models.
The approximate mean clump counts 
for the DSL model were $E[K] \approx 1.6,2.7,4.5$ 
for the three flow rates, $\lambda=0.1,0.2,0.3$, respectively.
The simulation experiment then had 
a crossed $3 \times 2 \times 3$ design, with $n=500$ independent
datasets generated per combination of $\lambda,t$ and $\sigma$.  
Normal plots and
Kolmogorov-Smirnov statistics did not indicate any obvious
non-normality for either the MLE or $\tilde\lambda$.



Table 1 summarizes the results of the simulation.   The bias of the
$M$-estimate relative to $\lambda$ and the efficiency relative to the
MLE are given in the middle section.
Though the bias of the MLE formulated under the DSL model dissipates
with increasing $\lambda$ or $t$,  it does not exhibit robustness
to heterogenous segment lengths, in the sense that it has larger variance than
the $M$-estimate. 
%
%
Empirical coverage probabilities for $95\%$ confidence 
intervals based on the LRT and those of the form $\tilde\lambda \pm 1.96 SE$ 
where $SE$ denotes the appropriate estimated asymptotic standard error
from Section 2 
are given in the right section of Table 1. 
For the shorter simulations $(t=1000)$, there is a tendency for coverage
probabilities based on $\tilde\lambda$ to be low. 
For datasets with a larger number of clumps ($t=10000$),
the nominal coverages for intervals based on $\tilde\lambda$ are 
reached.  With $n=500$ simulations, the Monte Carlo standard error is
such that any sample proportion less than 0.934 is significantly less
than the nominal 0.95 with comparisonwise error rate 0.05.
%
%
%
Additionally, the intervals around the $M$-estimate that
use the standard error, $SE_G$, which is a function of the sample
variance of the clump-lengths, appear to do better for the RSL models
with large $N(t)$, particularly for the noisy segment length
$\sigma=1$ case.
The likelihood ratio interval gives coverages consistent
with nominal levels in simulations with the DSL model, 
but breaks down under 
the RSL model where the likelihood is
misspecified. 
In summary, the recommendation based on these simulations is that
the $M$-estimator is reasonably efficient under the DSL model
and robust to the conditions of the RSL model.  Confidence intervals 
based on the standard error $SE_G$ meet nominal coverage
probabilities in large samples under either model.


\begin{table}
\caption{Simulation: relative efficiency and coverage probability of $\lambda$ estimators}
\centering
\begin{tabular}{ccc c cc ccc}  \hline
\multicolumn{3}{c}{Parameters} & & Rel. & Rel. & \multicolumn{3}{c}{Coverage probabilities}  \\
$\sigma$ & $t$ & $\lambda$ &  $\overline{N(t)}$ & Bias & Eff. & LRT & $SE(\tilde\lambda)$ & $SE_G(\tilde\lambda)$ \\ \hline
0  &  1000  &  0.1  &  60.1  &  0.01  &  0.89  &  0.966  &  0.950  &  0.944 \\ 
0  &  1000  &  0.2  &  73.2  &  -0.01  &  0.95  &  0.956  &  0.956  &  0.934 \\ 
0  &  1000  &  0.3  &  66.7  &  -0.01  &  0.98  &  0.948  &  0.946  &  0.944 \\ 
0  &  10000  &  0.1  &  605.8  &  0.00  &  0.83  &  0.940  &  0.942  &  0.942 \\ 
0  &  10000  &  0.2  &  734.5  &  0.00  &  0.88  &  0.960  &  0.956  &  0.950 \\ 
0  &  10000  &  0.3  &  669.6  &  0.00  &  0.97  &  0.942  &  0.938  &  0.942 \\  \hline
0.5  &  1000  &  0.1  &  60.2  &  0.56  &  8.3  &  0.128  &  0.922  &  0.916 \\ 
0.5  &  1000  &  0.2  &  73.0  &  0.15  &  2.2  &  0.768  &  0.910  &  0.902 \\ 
0.5  &  1000  &  0.3  &  66.3  &  0.05  &  1.1  &  0.940  &  0.952  &  0.950 \\ 
0.5  &  10000  &  0.1  &  606.1  &  0.57  &  82.4  &  0.000  &  0.942  &  0.946 \\ 
0.5  &  10000  &  0.2  &  736.1  &  0.15  &  17.2  &  0.002  &  0.946  &  0.954 \\ 
0.5  &  10000  &  0.3  &  670.1  &  0.05  &  3.3  &  0.646  &  0.938  &  0.940\\  \hline
1  &  1000  &  0.1  &  60.5  &  0.56  &  7.3  &  0.136  &  0.914  &  0.940\\ 
1  &  1000  &  0.2  &  73.5  &  0.13  &  1.9  &  0.798  &  0.922  &  0.920\\ 
1  &  1000  &  0.3  &  66.9  &  0.03  &  0.98  &  0.952  &  0.944  &  0.942 \\ 
1  &  10000  &  0.1  &  605.3  &  0.56  &  71.9  &  0.000  &  0.922  &  0.952 \\ 
1  &  10000  &  0.2  &  735.7  &  0.14  &  13.6  &  0.016  &  0.924  &  0.946 \\ 
1  &  10000  &  0.3  &  668.6  &  0.04  &  3.0  &  0.694  &  0.944  &  0.954 \\  \hline
\end{tabular}
\end{table}


\section{Estimation of total particle flow}

In the case where either $\lambda$ is known or variance in its estimation
is negligible,
total particle flow may be estimated by $E[A(t)]=\lambda t$.
When $t$ is not available, another estimator
can be formed by substitution of $t \approx \sum Y_i + \sum E(Z_i)$
into the expression giving $\widehat{E[A(t)]} = \lambda\sum Y_i + N(t).$

In the DSL model, clump orders ($K_1,K_2,\ldots$) may be 
shown (\cite{Pippenger}) to be 
geometrically distributed.
A clump is of order one ($K_i=1$) if there are
no arrivals within $t_0$ time units of the start of the clump, which occurs
with probability $\emla$.  A clump is of order two if there is
exactly 1 arrival within $t_0$ units and none 
in the next $t_0$ time units,
an event which
occurs with probability $(1-\emla)\emla$ and so on.
$K_1,K_2,\ldots$ are then independent geometric random variables with 
support on positive integers:
$$ \Pr(K_i=k) = (1-\emla)^{k-1}\emla \ \ \ \mbox{ for } k=1,2,\ldots$$
with $E(K_i)=e^{\lambda t_0}$ and $\Var(K_i)=e^{2\lambda t_0}-e^{\lambda t_0}$.
If the system is vacant when observation ends at time $t$, then
total particle flow may be expressed as the sum of these clump orders:
$ A(t)=K_1 + \cdots + K_{N(t)}$.  If the system is occupied at time $t$,
there is a partial clump that contributes a relatively
small amount of particle flow for large $t$.  Expressing total particle flow 
$A(t)$ as the sum of clump orders each with mean $e^{\lambda t_0}$ suggests the
estimator $\hat{A}_1(t;\lambda)=N(t)e^{\lambda t_0}$.
When evaluated at the $M$-estimator $\tilde\lambda$, with mean passage time 
$\mu=t_0$, the two estimators
of total particle flow
become equivalent: 
$\tilde{A}_0(\tilde\lambda) = N(t)e^{\tilde\lambda \mu} = \hat{A}_1(t;\tilde\lambda)$.

More efficiency might be gained by conditioning on the 
clump lengths.
The estimator $p(y)$ of an individual clump order which 
is a function of the clump length $y$ and minimizes the mean 
squared error $E[(K-p(y))^2]$, is the {\em Bayes}
estimate, or 
$p(y)=E(K|Y=y)$.  An estimate of mean total particle flow
$E[A(t)]=E[\sum K_i]$ is then given by
summing over clumps:
$$ \hat{A}_{B}(t;\lambda) = \sum_{i=1}^{N(t)} E(K_i|Y_i;\lambda). $$
Of course $E[K_i|Y_i=t_0;\lambda]=1.$  
The approach used by \citet{hall1} to derive the clump length density
$f(y)$ in the DSL model may be extended to obtain the  conditional
mean of clump orders, $E(K|Y)$.
%
%
Let the beginning of a clump be the origin and let $k$ denote
an integer greater than unity.
The joint event $K=k$ and
$Y\in (y,y+dy)$ 
occurs if and only if there is 
a particle arrival at $(y-t_0,y-t_0+\Delta y)$, no arrival 
in $(y-t_0+\Delta y,y)$, exactly $k-2$ arrivals in $(0,y-t_0)$, and the nearest 
neighbor of each of these $k-2$ arrival times is not further than $t_0$ time 
units away.  Since the first three of these conditions are independent
and the fourth is conditionally independent of the first two given the
third, the joint probability of these four events is the product
$$\lambda \Delta y \emla \frac{(\lambda(y-t_0))^{k-2}}{(k-2)!}e^{-\lambda(y-t_0)} p_{k-2}\left(\frac{t_0}{y-t_0}\right)$$ where $p_n(u)$ denotes the chance 
that the largest division formed by a random sample of $n$ points taken
from the unit interval does not exceed $u$.   This probability is given by 
\begin{eqnarray*}
p_n(u) 
& = & \sum_{j=0}^{[u^{-1}]} (-1)^j \choose{n+1}{j} (1-ju)^n \\
& = & 1 - (n+1)(1-u)^n + \choose{n+1}{2}(1-2u)^n - \ldots
\end{eqnarray*}
where $[\cdot]$ denotes the largest integer not exceeding the argument.
Division by $f(y)$ and differentiation with respect to $y$ yields the 
conditional density 
$$ \Pr(K=k|Y=y) = \frac{\lambda e^{-\lambda y}}{f(y)}\frac{(\lambda(y-t_0))^{k-2}}{(k-2)!} p_{k-2}\left(\frac{t_0}{y-t_0}\right). $$
If $s=[y/t_0]$, then summation over positive integers yields an exact 
expression for the conditional mean:
\begin{eqnarray*}
E(K|Y=y) & = & \sum_{k=s+1}^\infty k \Pr(K=k|Y=y) \\
& = & \frac{\lambda e^{-\lambda t_0}}{f(y)}\sum_{k=s+1}^\infty k \frac{\left(\lambda(y-t_0)\right)^{k-2}}{(k-2)!}p_{k-2}(\frac{t_0}{y-t_0}) \\
& = & \frac{\lambda e^{-\lambda t_0}}{f(y)}\sum_{k=s+1}^\infty k \frac{\left(\lambda(y-t_0)\right)^{k-2}}{(k-2)!} \sum_{j=0}^{s-1} (-1)^j \choose{k-1}{j}\left(1-\frac{jt_0}{y-t_0}\right)^{k-2}.
\end{eqnarray*}

Inspection of $\Pr(K=k|Y=y;\lambda)$ reveals 
that for $t_0<y<2t_0$, 
$K$ has the translated Poisson distribution
with mean and variance that are linear in $y$.
For larger $y$, 
numerical evaluation of $E(K|Y=y)$ be 
difficult.
%
%
Inspection of plots for larger $y$ and various values of $\lambda$
indicates that after a jump discontinuity 
of $\lambda t_0 \emla (1-\emla)^{-1}$ 
at $y=2t_0$,
approximate linearity extends to $y>2t_0$.
For cases where $N(t)$ is large and there is heavy
clumping, $E(K|Y=y)$ can be approximated by linear interpolation to save 
computational effort.

\subsection{Simulation}

The performances of these estimators of mean total particle
flow are compared 
using the simulated data from section 2.  
Error for either $\hat{A}_1$ or $\hat{A}_B$, as a percentage of 
the mean particle flow 
is assessed using 
the relative root mean squared error, RRMSE:
$$ RRMSE(\hat{A}(t)) = \frac{1}{\overline{A(t)}} \sqrt{500^{-1}\sum_i (\hat{A}_i(t)-A_i(t))^2}$$
where $i$ indexes the 500 simulated datasets.
Table 2 summarizes relative bias and RRMSE 
of estimates obtained by substitution of
the $M-$estimates $\tilde\lambda$ into the expressions $\hat{A}_1(t)=N(t)e^{\lambda \mu_t}$ and $\hat{A}_B(t;\lambda)$
for each
simulated experimental condition.
The estimation based on clumpwise estimated clump orders 
$\hat{A}_B$, is competitive under the DSL model ($\sigma_t=0$) 
for smaller sample sizes, ($t=1000$).  It suffers from some positive
bias in RSL models that appears to decrease as flow rate $\lambda$ increases,
though it remains inferior to $\hat{A}_1$
despite smaller variance and higher correlation with $A(t)$. 
In the RSL model, many singleton clumps
have clump lengths slightly in excess of the mean singleton passage 
time $\mu_t$ and so 
have
estimated orders in excess of 1.  This may lead to a positive bias
for the clumpwise estimators which is particularly acute
when support is high near $Y=\mu_t$.  This theory is supported by the
poor performance under light clumping, when $\lambda=0.1$ and density
near $Y=\mu_t$ is highest among values of $\lambda$ considered in the
simulation.

\begin{table}
\caption{Error in estimation of total particle flow from simulations.} 
\centering
\begin{tabular}{ccc cc cc} 
\multicolumn{3}{c}{Parameters} & \multicolumn{2}{c}{Relative bias} & \multicolumn{2}{c}{RRMSE} \\
$\sigma_t$ & $t$ & $\lambda$ & $\hat{A}_1$ & $\hat{A}_B$  &$\hat{A}_1$ & $\hat{A}_B$ \\ \hline
0  &  1000  &  0.1  &  -0.008  &  -0.007  &  0.042  &  0.033 \\ 
0  &  1000  &  0.2  &  -0.012  &  -0.011  &  0.048  &  0.045 \\ 
0  &  1000  &  0.3  &  -0.014  &  -0.013  &  0.049  &  0.048 \\ 
0  &  10000  &  0.1  &  -0.001  &  -0.001  &  0.014  &  0.01 \\ 
0  &  10000  &  0.2  &  0.000  &  0.000  &  0.015  &  0.014 \\ 
0  &  10000  &  0.3  &  -0.002  &  -0.002  &  0.016  &  0.015 \\  \hline
0.5  &  1000  &  0.1  &  -0.003  &  0.178  &  0.048  &  0.185 \\ 
0.5  &  1000  &  0.2  &  -0.008  &  0.061  &  0.051  &  0.076 \\ 
0.5  &  1000  &  0.3  &  -0.011  &  0.013  &  0.054  &  0.052 \\ 
0.5  &  10000  &  0.1  &  0.000  &  0.184  &  0.014  &  0.184 \\ 
0.5  &  10000  &  0.2  &  -0.001  &  0.067  &  0.015  &  0.068 \\ 
0.5  &  10000  &  0.3  &  -0.003  &  0.022  &  0.015  &  0.026 \\  \hline
1  &  1000  &  0.1  &  -0.002  &  0.189  &  0.058  &  0.198 \\ 
1  &  1000  &  0.2  &  -0.010  &  0.063  &  0.053  &  0.079 \\ 
1  &  1000  &  0.3  &  -0.011  &  0.018  &  0.052  &  0.052 \\ 
1  &  10000  &  0.1  &  0.001  &  0.192  &  0.018  &  0.193 \\ 
1  &  10000  &  0.2  &  0.000  &  0.072  &  0.017  &  0.074 \\ 
1  &  10000  &  0.3  &  -0.001  &  0.026  &  0.016  &  0.031 \\  \hline
\end{tabular}
\end{table}


In summary, for minimal relative error, these simulations suggest
the use of the simple $\hat{A}_1(t)$ estimator, which is 
unbiased and involves less computation than the clumpwise
estimator $\hat{A}_B(t)$.  A slight loss of efficiency
under the DSL model may be offset by the superior performance in the RSL model.
Expressed relative to total particle flow, the root MSE was not larger than
$5.8\%$ in any of the conditions simulated here.

\section{Experimental data}

An optical sensor was used to measure clump lengths and clump 
velocities in experiments 
(\cite{grift2,grift3}) 
in which a known number of spherical
particles was 
dropped through a device simulating an aerial 
spreader duct.
Various quantities of several kinds of particles (BBs, urea fertilizer) 
were dropped at several velocities.  The data considered here
include 10 runs with 4000 identical steel particles (BBs) 
dropped from each of two heights and 5 runs with 2000 BBs dropped 
from a fixed height.  
Mean ($\bar{y}$) and variance ($s_y^2$) of physical 
lengths (in $mm$) appear in Table 3 along with
other statistics from the experiments.
%
Division by mean velocity ($\bar{v}=2.23 mm/msec$) was used to 
transform the measurements to the time line (in $msec$) to obtain Figure 1. 
In general, velocity was reasonably constant within a run of the
experiment. 

The data were imperfect and some outlier removal was undertaken.
For example, the counter returned several clumps with negative 
velocities or 
negative physical lengths, or sometimes both.  Additionally, each run 
contained a very small number of extremely short clumps, much less than the
particle diameter, possibly due to matter other than the particles of 
interest blocking the sensor.  The number of questionable clump 
measurements that were removed did not exceed $1\%$ for any of 
the 25 runs.


\begin{table}
\caption{Estimation from experiments with BBs}
\centering
\begin{tabular}{c ccc cccc } \hline
Run & $N$ & $\bar{y}$ & $s_y^2$ & $\tilde\lambda(SE)$ & $\hat{A}_1$ & $\hat{A}_B$ \\ \hline
1  &  2958  &  5.22  &  3.07  &  0.070 (0.003)  &  4041  &  4921 \\ 
2  &  2930  &  5.22  &  2.91  &  0.070 (0.003)  &  3997  &  4946 \\ 
3  &  2891  &  5.26  &  3.39  &  0.073 (0.003)  &  4008  &  4874 \\ 
4  &  2935  &  5.22  &  3.00  &  0.070 (0.003)  &  4000  &  4944 \\ 
5  &  2990  &  5.16  &  2.88  &  0.065 (0.003)  &  3986  &  4941 \\ 
6  &  2941  &  5.20  &  3.00  &  0.068 (0.003)  &  3984  &  4883 \\ 
7  &  2983  &  5.15  &  2.84  &  0.064 (0.003)  &  3969  &  4900 \\ 
8  &  2956  &  5.16  &  2.90  &  0.065 (0.003)  &  3952  &  4846 \\ 
9  &  2894  &  5.24  &  3.12  &  0.071 (0.003)  &  3976  &  4831 \\
10  &  2914  &  5.25  &  3.11  &  0.073 (0.003)  &  4025  &  4931 \\ \hline
11  &  1821  &  6.76  &  11.77  &  0.176 (0.005)  &  3988  &  4299 \\  
12  &  1770  &  6.85  &  11.56  &  0.182 (0.005)  &  3976  &  4303 \\ 
13  &  1805  &  6.80  &  12.30  &  0.179 (0.005)  &  4000  &  4321 \\ 
14  &  1748  &  6.96  &  12.57  &  0.188 (0.005)  &  4038  &  4333 \\ 
15  &  1800  &  6.85  &  12.13  &  0.182 (0.005)  &  4040  &  4340 \\ 
16  &  1784  &  6.93  &  14.82  &  0.186 (0.005)  &  4089  &  4403 \\ 
17  &  1772  &  6.93  &  12.56  &  0.187 (0.005)  &  4064  &  4341 \\ 
18  &  1788  &  6.89  &  13.28  &  0.184 (0.005)  &  4052  &  4346 \\ 
19  &  1812  &  6.78  &  12.01  &  0.178 (0.005)  &  3995  &  4317 \\ 
20  &  1790  &  6.84  &  11.98  &  0.181 (0.005)  &  4005  &  4330 \\ \hline
21  &  746  &  7.54  &  17.10  &  0.219 (0.008)  &  1981  &  2143 \\  
22  &  791  &  7.24  &  13.20  &  0.204 (0.007)  &  1959  &  2141 \\ 
23  &  777  &  7.46  &  13.15  &  0.215 (0.007)  &  2024  &  2184 \\ 
24  &  774  &  7.30  &  13.23  &  0.207 (0.007)  &  1941  &  2102 \\ 
25  &  745  &  7.57  &  13.39  &  0.221 (0.007)  &  1989  &  2133 \\ \hline
\end{tabular}
\end{table}

In these experiments, total particle flow is fixed and total flow 
time varies with run 
and is not observed.  The opposite 
is true for the application of mass flow measurement during aerial 
application of fertilizer particles.  The theoretical results regarding
inference for the random particle flow $A(t)$ for fixed $t$ 
do not necessarily hold under
the conditions of the experiment, where $A(t)$ is fixed and
$t$ varies and is not observed.
However, Table 3 provides some indication that estimates for 
total particle flow, $A(t)$, have good empirical performance when 
it is treated as random, at least in these experiments.


The observed value of the estimator $\hat{A}_1$ is given in the penultimate
column.  It appears to perform reasonably well 
under these conditions. 
The average of $\hat{A}_1$ over runs 1-20 is 9 and the root
mean squared error from 4000 is 35.5, which is 0.9\% of the target.
There is some evidence of positive bias in the 
high intensity runs 11-20.  
A two-sided $t$-test of the hypothesis that $E[\hat{A}_1]=4000$
under the conditions of runs 11-20 yielded a $p$-value of $0.065$ on $df=9$.

Higher flow rates lead to more clumps per particle, fewer singletons,
and larger variance in estimation of clump order, either conditionally
as in $\hat{A}_B$ or unconditionally, as in $\hat{A}_1$. 
The standard deviations of $\hat{A}_1$ under the light (runs 1-10) and heavy
(runs 11-20) clumping 
conditions with 4000 BBs were $s_l=26.4$ and
$s_h=37.1$, respectively.
The estimates $\hat{A}_B$, which are based upon the DSL model,
exhibit substantial positive bias, as they did for 
data simulated
under the RSL model.  The same is true for the MLE of $\lambda$.

To assess the goodness of fit of the linear Boolean models, probability
histograms of the clump length data were 
checked for agreement with the 
estimated density $f(y;\tilde\lambda,d)$.  One such check appears in
Figure 1, which exhibits reasonable fit except for slightly 
lowered mass at the mean segment length, $\mu=2 msec$ and slightly
more observations just above the mean segment length than expected
under uniformity of this part of the density.  All of the other histograms
exhibited the same three distinctive features of a spike
near this fixed segment length, near uniformity between one and two
of these lengths and a long right tail.
Quantile plots and Kolmogorov-Smirnov goodness-of-fit tests,
for estimates in runs 1-10 or runs 11-20 
do not indicate any non-normality 
in the distribution of $N$, $\tilde\lambda$ or $\hat{A}_1$.

\section{Conclusion}
Two versions of a simple linear Boolean model are proposed to describe
passage times of clumps of particles in a type II counter system; one
assumes deterministically equal passage times for all particles, while
the other assumes these to be distributed about a known mean with unknown
variance.  An $M$-estimator of flow intensity 
is developed that is intuitively 
sensible, computationally feasible, and robust to conditions 
where either particle velocity and/or diameters have substantial 
variability or are being measured with error by the type II counter.
%


For total mass flow, $A(t)$, two estimators are developed.
The first is simply product of the number of clumps, $N(t)$ and the
estimate of the mean number of particles per clump.  The second more
complex estimator is the clumpwise sum of conditional mean clump orders ($K$),
given clumplengths ($Y$).
%
In models where segment lengths were deterministically equal,
the Bayes estimator exploiting the conditional mean clump order 
had relative root mean squared error not exceeding
$5.0\%$, and always lower than that of the simpler estimator
based only on the $M$-estimate of flow rate and the number of clumps.
Under the most favorable conditions, with light particle flow
and a long dispersal period, the relative error was as small
as $1\%$.
%
%
While the Bayes estimator did well in data simulated from the DSL
model, it was outperformed by the simpler estimator in simulations
where the segment lengths vary according to a normal distribution
and in the bench-scale experiments.
The relative root mean square error when using the 
estimator of based
ranged between $1.4\%$ and  $5.1\%$. 
The relative
root mean square errors for the experimental 
data were $0.6\%$ and $1.1\%$, in the low and high intensity 
runs with 4000 BBs, respectively and $1.8\%$ in the runs
with 2000 BBs.  

\bibliographystyle{chicago}
\bibliography{refs-short}
\end{document}